# SecDOAR: A Software Reference Architecture for Security Data Orchestration, Analysis and Reporting

Muhammad Aufeef Chauhan[a, b], Muhammad Ali Babar[a, b], Fethi Rabhi[c]

[a]*Cyber Security Cooperative Research Centre, Australia*

[b]*School of Computer Science, The University of Adelaide, Adelaide, 5005, South Australia, Australia*

[c]*School of Computer Science and Engineering, University of New South Wales, Sydney, 2052, New South Wales, Australia*

**Abstract**

A Software Reference Architecture (SRA) is a useful tool for standardising existing architectures in a specific domain and facilitating concrete architecture design, development and evaluation by instantiating SRA and using SRA as a benchmark for the development of new systems. In this paper, we have presented an SRA for Security Data Orchestration, Analysis and Reporting (SecDOAR) to provide standardisation of security data platforms that can facilitate the integration of security orchestration, analysis and reporting tools for security data. The SecDOAR SRA has been designed by leveraging existing scientific literature and security data standards. We have documented SecDOAR SRA in terms of design methodology, meta-models to relate to different concepts in the security data architecture, and details on different elements and components of the SRA. We have evaluated SecDOAR SRA for its effectiveness and completeness by comparing it with existing commercial solutions. We have demonstrated the feasibility of the proposed SecDOAR SRA by instantiating it as a prototype platform to support security orchestration, analysis and reporting for a selected set of tools. The proposed SecDOAR SRA consists of meta-models for security data, security events and security data management processes as well as security metrics and corresponding measurement schemes, a security data integration model, and a description of SecDOAR SRA components. The proposed SecDOAR SRA can be used by researchers and practitioners as a structured approach for designing and implementing cybersecurity monitoring, analysis and reporting systems in various domains.

*Keywords:* Security Reference Architecture, Security Data Orchestration, Security Data Analysis, Security Data Reporting, Security Metrics, Security Meta-models.

## 1. Introduction

Security Data Orchestration, Analysis and Reporting (SecDOAR) encompass a number of aspects including tools for gathering and generating security data, metrics for analysing security data, generating events corresponding to the analysed data, and generating, analysing and validating the results for mitigation strategies [1]. For incorporating the aforementioned aspects of security data in security monitoring infrastructure, the architecture of the infrastructure should be designed to cater to SecDOAR needs. For example, to support integration among the security data, syntactic interoperability techniques such as the use of ontology along with a Service Oriented Architecture (SOA) can be adopted. Alternatively, adopting data exchange or data interoperability standards for relating different security data concepts can be used to integrate security data.

Security Information and Event Management (SIEM) tools require studying threat intelligence mechanisms, integrating threat intelligence with security data architecture and incorporating processes for effectively reducing the business risk of threats to security data [2]. A threat intelligence mechanism should be added to the system in a way that enhances security awareness capability without having a significant impact on existing organisational processes. Consumable threat intelligence is used for both initial threat detection as well as to determine the context around the detection of threats [2]. Events associated with threat intelligence need to be filtered as well before these can be processed.

Interoperability of the security data in SIEM tools is vital to providing an end-to-end cybersecurity monitoring solution because a single security tool cannot cover all aspects of SecDOAR [3]. For supporting interoperability among the security tools, different security standards need to be incorporated to organise and manage security data. The following aspects related to security data need to be explored for this purpose. (i) What standards apply to security data in a particular domain? (ii) What is the nature of the standards? (iii) At which level of data security, a particular security standard can be applied? For example, each of the network data security, transactional data security and enterprise logs security may require different security standards for analysis. Moreover, the boundaries of the security data standards need to be established to determine to which aspect of the system a particular standard can be applied. For example, there can be different standards and guidelines for managing the heterogeneity of data sources, there can be limitations associated with the application of a particular standard in

*Email addresses:* aufeef.chauhan@adelaide.edu.au (Muhammad Aufeef Chauhan[a, b]), ali.babar@adelaide.edu.au (Muhammad Ali Babar[a, b]), f.rabhi@unsw.edu.au (Fethi Rabhi[c])

a specific domain, or a security standard can be only applicable at a specific application level.

Research and development of security monitoring, analysis and reporting tools focus on different aspects of cyber security. However, current security information and event management tools are limited in terms of their security orchestration, analysis and reporting capabilities [4]. While these tools typically provide features to provide correlation between logs from firewalls, domain controls and failed connections, these tools cannot deal with high-level security risk metrics that can provide quantitative evidence [4]. There is also a need to support event-based anomaly detection in addition to heuristics and signature-based approaches [4]. Therefore, to have a comprehensive and consolidated overview of the cyber security status of an enterprise system, system design guidelines are required. An SRA is a well-established methodology for providing architecture-centric guidelines and solutions either by proposing a new solution or by standardising existing solutions [5][6].

In this paper, we have provided a high-level security data-centric SRA to support Security Data Orchestration, Analysis and Reporting (SecDOAR). The focus of this paper has been to provide an integrated SecDOAR architecture. We have used existing literature on security data standards and security data metrics as a reference to guide the design and development of SesDOAR SRA. The SecDOAR SRA presented in this paper consists of meta-models to represent different aspects of the security data, metrics to capture and analyse the security data, a semantic integration model for security data integration, and SecDOAR SRA elements (layers, components and relations among them). We have documented and elaborated SecDOAR SRA in terms of its context, goals, design strategy, evaluation and instantiation guidelines [7][8]. The results of the evaluation and prototype implementation demonstrate that the proposed SecDOAR SRA can be used to provide an end-to-end cybersecurity solution to capture, comprehend, perceive, analyse and report security data.

The research work presented in this paper makes the following key contributions:

- We have presented meta-models for security data, security events and security data management processes. A security data meta-model identifies different elements of security data architecture and relations among the elements. A security events meta-model can be used to identify events associated with the identification of security threats and corresponding remedies. A security data management process meta-model describes different phases for capturing, processing and interpreting security data as well as generating responses. We have also elaborated on the process for security data understanding and management.

- We have synthesised metrics that are used to capture security data. Each metric has been elaborated in terms of what part of security data can be captured by the metric, what measurement schemes need to be incorporated to capture relevant security data, and what security data-related analysis can be performed using the metric.

- We have proposed a semantic model that shows relations between different elements of the security data and a description language to facilitate the composition of different components and security tools to support SecDOAR operations.

- We have presented layers and components of SecDOAR SRA. We have demonstrated the feasibility of the presented SecDOAR SRA by implementing a prototype platform to support SecDOAR operations for Denial of Service (DoS) and Distributed Denials of Service (DDoS) attacks. We have also evaluated SecDOAR SRA by providing a comparative analysis of the presented SRA with existing commercial SIEM tools and platforms.

This paper is organised as follows. Section 2 provides an overview of the existing solutions for software security information and event management. Section 3 describes the scope of the presented SRA and associated design methodology. Section 4 describes security data standards and key security recommendations corresponding to the security standards. Section 5 provides meta-models for security data, security events and security data management processes. Section 6 presents metrics that can be used for security data intelligence, information capturing and sharing. Section 7 explains the semantic model for explaining relations among elements along with a formalisation of the SecDOAR SRA composition and section 8 explains components of the presented SecDOAR SRA. Section 9 provides an evaluation of the SecDOAR SRA. Section 10 concludes the paper and provides direction for future work.

## 2. Dimensions of Security Data Orchestration, Analysis and Reporting Architectures

We have classified the existing work on security data and security architectures in the following subsections. The first subsection covers security reference architectures and concrete architectures for security-focused distributed and cloud-based systems. The second subsection covers solutions describing security data orchestration, analysis and reporting. We have used the concepts discussed in the existing work to derive elements of the SecDOAR SRA.

*2.1. Security Architectures*

A number of architectures and security solutions have been proposed to achieve specific aspects of software security. A generic and abstract reference architecture solution has been proposed by NIST for the security of cloud-based systems [9]. The NIST reference architecture identifies different layers and subsystems to handle security in a cloud environment. The proposed NIST reference architecture includes components for secure cloud service management, secure interoperability, secure provisioning, secure business support, secure deployment and cloud auditing [9]. Australian government information security manual has identified key elements of cyber security event logging and monitoring systems [10]. The guidelines advise capturing sufficient details for the event data to be useful. The guidelines also advise having a centralised event logging system that can capture and manage events from different monitoring systems, and maintain the history of event logs.

The reported concrete architecture focuses on achieving security in terms of authentication, multi-tenancy management, resource monitoring, resource scheduling, interoperability, service discovery and scalability to ensure the availability of the hosted data and services [11]. Fernandez et al. [12] have also proposed a security reference architecture for cloud computing environments. The authors have described security vulnerability and best practices for mitigating the vulnerabilities in cloud computing environments, however, their work does not provide insight into security from cloud environment utilization and application development perspective.

Several studies have reported research identifying specific security challenges that can be exploited by attackers as well as solutions to the reported challenges. Daniel [13] has presented data privacy issues and recommended practices. The most comprehensive work in this regard is conducted by Fernandes et al. [14] and Huang et al. [15] in which the authors have surveyed the challenges associated with public IaaS clouds and have shed some light on the practices that can be adopted to counter the threats.

Orchestration of the security platforms aiming for integrated security solutions has also been a focus of research in recent years. The work from Islam et al. [1] has provided a comprehensive review of the research on the security orchestration domain. The authors have described that security orchestration research has focused on unifying security tools, translating complex security processes of organisations into streamlined workflows and automatic generation of security incident response plans. Security orchestration can encompass unifying security tools and presenting results from different tools, performing threat intelligence based on the data, and automatically applying security orchestration. Two key limitations of the existing work in the security orchestration domain have been identified as lack of interoperability among the tools, and lack of support for collaboration and coordination among the tools [1]. In their subsequent work, Islam et al. [16] have presented a high-level architecture for SIEM tools' application programmable interfaces (APIs) and have identified different layers including integration, data processing, semantic and orchestration layers.

*2.2. Design of Security Data Orchestration, Analysis and Reporting Systems*

Several architectural centric solutions have been proposed for supporting SecDOAR operations. Semantic integration is an approach to structure heterogeneous data originating from multiple sources in a way that semantic relationships among related data elements can be established [17]. Common elements for structuring security data to support data interoperability include security information database, ontology (or semantic integration models) database, ontology structure, ontology reasoning and data collection elements [17].

Extraction, conversion, cleaning and loading are four stages of security data preprocessing. In addition, temporal characteristics are important while extracting and managing security data. Multiple steps can be taken to safely share the security data [18]. The first step is to select data from sources based on security requirements, legal constraints, user data sharing consent and security data interoperability potential. The second step is to access data from data stores or databases. The third step is to provision data based upon data usage requirements and agreements. In the final step, semantically integrated data is shared among multiple collaborating systems. The systems supporting integration can employ a number of techniques including text mining, data transformation and harmonisation, privacy and security risk management, and shared resources and registries [18].

Combining event-driven systems with the processing of the data from heterogeneous data sources requires a semantic parser that can be used to relate elements of security data with elements of event logs [19]. The semantic parser can be used in combination with security data events meta-model (discussed in Section 5.2). The semantic parser should consider hierarchical relations among the entities [20].

*2.2.1. Security Data Orchestration*

A number of papers have identified different elements for security data orchestration and how it can be used to complement security analysis and reporting. The security orchestration framework can consists of a number of layers for security orchestration [21]. The application layer can manage security services and how the security services interact with each other. The physical layer can control physical devices and system components. The data-driven task instructions can help to perform security orchestration in accordance with the requirements of the security data under examination. Scheduling orchestration instruction using historical data knowledge and issuing commands and instruction to security orchestration components accordingly is important as well.

A loosely coupled approach labelled as Service Oriented Software Defined Security (SOSDSec) framework that separates security management from underlying security controls is presented in [22]. The approach abstracts security operations into generic reusable security services. The framework emphasises on the following key elements for security orchestration. (i) Security policy model for managing policies for different security assets. (ii) Managing capability of security controls. (iii) Managing security control capability requirements and respective descriptions. (iv) Ontology-based model for matching requirements of the security controls with capabilities of the security controls. (v) Mapping security data services to the corresponding assets. (vi) Providing interfaces through which security services can receive commands for capturing security data. (vii) Explicit interfaces to receive security events associated with security data. (viii) Execution of security orchestration plans corresponding to security events under investigation.

An architecture for provisioning security services and managing security data at edge nodes have been proposed in [23]. This work has proposed a service life cycle management engine, which is responsible for launching, monitoring and terminating the services for capturing security data. A monitoring engine has been proposed to monitor security data corresponding to the required metrics. A data-driven message oriented middleware consisting of an orchestration engine for network anomaly detection has been presented in [24]. The anomaly detection module detects anomalies corresponding to the desired metrics. A module for elastic and scalable data collection and search is needed for handling large values of input security data. The process for anomaly detection consists of three phases: training phase (read data and triggers corresponding anomaly detection module), detection phase (monitor data collection channels and perform analysis on the data, send detection results to security analyser, if an anomaly is detected then prepare and write anomaly data for search), orchestration phase (resource scheduling and corresponding and synthesising security data according to the required security policies).

A layered architecture for security orchestration policy, monitoring and secured computing is proposed in [25]. A security orchestration policy manager manages policy decisions and security risks. The secured computing is done by processing a tenant's data in accordance with the tenant's security requirements. The orchestration process for security data is presented in [26]. The orchestration process assigns security tags to security data and services (to process the data). Three types of tags are added: trusted tag for highly secured data and services, semi-trusted tag for partially secured data and services, and public tag for data and services that are not secured. Security data is sent to services for processing based upon the assigned security tag.

An architecture for the management of the orchestration layer is presented in [27]. The architecture consists of a virtual infrastructure manager that manages underlying virtualised resources, application and service management component for managing services. An architecture for security data orchestration for cloud-based services is presented in [28]. A security broker pattern is used to manage access to the cloud-hosted data. An orchestration protocol for balancing security and performance is presented in [26].

Cyber threat intelligence (CTI) data can be gathered from several internal or external sources [29]. Internal threat sources include but are not limited to system logs, network events, network traffic and human sources [29]. External sources of threats include sources such as common vulnerability and exposure (CVE) and common weakness enumeration (CWE) [29]. External sources of threats expose data in specific standardised formats such as Structured Threat Information Expression (STIX) and Trusted Automated eXchange of Intelligent Information (TAXII).

### 2.2.2. Security Data Analysis

Scanning data sources based upon security instructions helps to identify threats and prepare appropriate response and reporting plan is important for security data analysis [21]. Validation of data for trustworthiness is critical as well [30].

### 2.2.3. Security Data Reporting

Threat intelligence building blocks are important for threat intelligence and reporting [31]. Testing of the security data with reference to security standards helps to identify threats [32]. The ontology-based thread modelling framework presented in [33] signifies the need for a structured approach for modelling threats that can be associated with security data. A knowledge graph for identifying security gaps and situational awareness is required for intelligent decision making. An intelligent security architecture for dynamic attack detection in accordance with updated security policies has been presented in [34]. The architecture uses a decision management engine for security analysis and reporting. An appropriate trust level can be associated with security data [35]. Additional security measures can be put in place with data including homomorphic encryption.

A privacy attack model and risk assessment technique based upon the attack tree is presented in [36]. Risk assessment techniques are based upon the degree of threat on the data. Prior knowledge of attacks can be used to prevent future attacks. Threat intelligence layer integration with security policy can be used for security orchestration analysis and reporting [37].

## 3. Research Motivation, SecDOAR SRA Design and Evaluation Approach

The primary goal of the research presented in this paper has been to provide an architecture-centric solution that can facilitate the integration of security data across multiple tools for security data orchestration, analysis and reporting. A security operation team in an organisation often uses multiple tools (or platforms) to monitor its Information Communication and Technology (ICT) infrastructure for identifying security threats and responding to security breaches. As each tool (or platform) focuses on different aspects of cybersecurity (e.g., network attacks or distributed denial of service attacks), monitoring the security of the whole ICT infrastructure for potential security breaches using a single tool provided by a specific vendor is not possible. Therefore, integration and interoperability of security data are needed to provide an end-to-end security solution. By providing SecDOAR SRA, we aim to guide the design and development of the integrated tolls (or platforms) for security data orchestration, analysis and reporting so that security operation teams can use suitable tools based on the available security data.

Figure 1 shows the overall process of security data orchestration, analysis and reporting that we have synthesised in SecDOAR SRA. The security data are captured or generated (by observing ICT infrastructure) by different tools, e.g., LimaCharlie [38], Wireshark [39], Splunk [40], Sonarqube [41], Zeek [42] and Intruder [43]. The data captured by each tool provide an insight into a particular aspect of cyber security. For example, Wireshark provides details on the data volume being received or sent out a physical server hosting applications. LimaCharlie can be used to capture and monitor the network traffic. Integration models or semantic meta-models are used to combine the data from multiple sources (i.e., tools) and provide a consolidated view of the data related to a particular aspect of the security. For example, to identify denial of service (DoS) or distributed denial of service (DDoS) attacks, information about IP addresses of sources of requests, destinations of requests, ports (or hosted application) at which the data is being received and the volume of data received over a particular time unit are some of the factors to consider that if a system in under DoS or DDoS attack. Security data analysis metrics can be used to analyse the consolidated security data and make inferences from the data. In the final stage, the results of the analysis can be synthesised to be sent to either Graphical User Interfaces (GUIs) to be used by security operation team members or external tools for automatically deploying mitigation strategies.

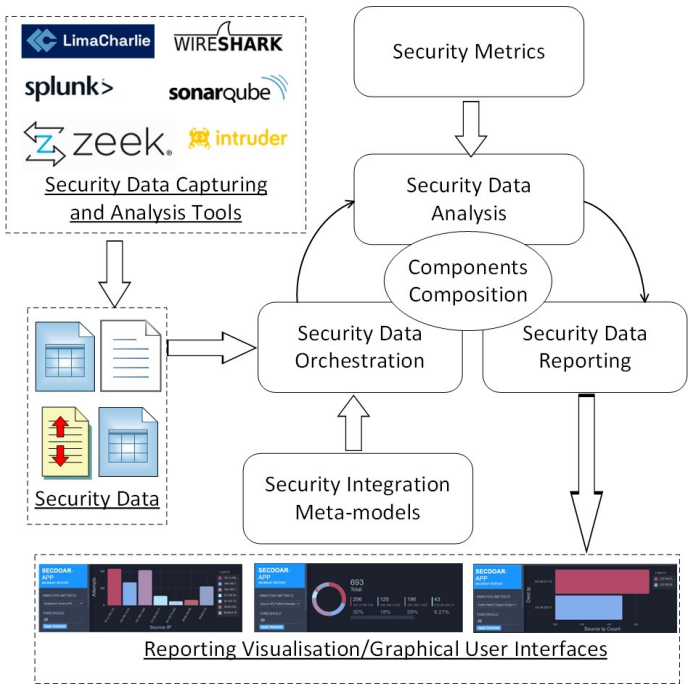

Figure 1: Security Data Orchestration, Analysis and Reporting Process

*3.1. Selection of Data Sources for Building the SecDOAR SRA*

Selection of sources of information through which the concepts and elements for an SRA can be derived is important. We extracted main components of the SecDOAR SRA from the concrete solutions proposed in the studies discussed in Section 2.

*3.2. Research Gap and Our Contributions by proposing SecDOAR SRA*

Several solutions have been proposed in the literature (as described in related work in Section 2) for incorporating security in distributed, cloud-based and stand-alone systems, and for supporting integration among different types of security tools to support security operations. We have leveraged concepts proposed in the existing literature to provide a holistic security data-centric SRA that focuses on security data orchestration, analysis and reporting layers.

*3.3. Documentation and Evaluation Approach*

We provide details on the SecDOAR SRA documentation strategy in terms of context, goals, maturity stage, design, evaluation and instantiation. Evaluation and instantiation of a SRA is important. A SRA is evaluated in terms of its completeness with reference to the objectives and whether or not it covers all the concepts for the concrete architectures in a domain [7] . A SRA is also evaluated by showing that it is implementable [9]. We have evaluated SecDOAR SRA by comparing it with given objectives, existing industry platforms and by implementing a prototype to orchestrate, analyse and report security data gathered using open source and commercial tools. The details of the SecDOAR design, documentation and evaluation are described in Table 1.

## 4. Security Data Standards and Models

A number of security data standards have been reported in the literature that elaborate key elements of the security corresponding to different layers of applications. Some of the standards are generic and can be applied to any domain, whereas others are specific to a particular domain. General security standards such as US digital security standard IRS P1075 [45] focus on security concern trends, data vulnerability, unenforceable controls, under-defined processes and ambiguous specifications. The frameworks reported in the literature for monitoring and standardising security compliance argue that evidence-gathering mechanisms for security compliance should include measurable security, safety and organisational indicators (or metrics) [46]. Measurable security indicators are proposed in the IEC 62443-3-3[1] security standard and include secure identification and authentication, the strength of password-based authentication, and concurrent session management controls. Measurable safety indicators are proposed in the IEC 61508-3[2] security standard that recommends having time-triggered architecture, techniques and measures for error detection, and a controlled systematic approach for handling operational failures. Measurable organisational indicators are provided in ISO/IEC TS 33052[3] security standard that recommends having indicators for event logging, restrictions on software installations, and controlled access to the network and network services.

In addition to security standards providing guidelines for enhancing security of the systems and organisations, application-specific standards have also been proposed such as Structured Threat Information eXpression (STIX) and Trusted Automated eXchange of Indicator Information (TAXII) [47]. The STIX standard was proposed for network security and threat intelligence expressions [48]. STIX provides a framework for expressing threat intelligence, improving the accuracy of threat intelligence and supporting interoperability with other threat intelligence systems. These features help to automate the threat intelligence process for analysing cyber threats, understanding patterns of threats, managing responses to the threats, and sharing threat information with other systems. The TAXII standard was proposed for supporting network security threat intelligence exchange mechanisms [47]. TAXII relies upon using HTTPS protocol for exchanging security intelligence. Threats intelligence can be organised using STIX protocol and shared with other systems using TAXII framework. A number of software architecture design methodologies including client-server or publisher-subscriber can be adopted for implementing information exchange using TAXII.

---

[1] https://webstore.iec.ch/publication/7033
[2] https://webstore.iec.ch/publication/5517
[3] https://www.iso.org/standard/55142.html

Table 1: SecDOAR SRA Design, Documentation and Evaluation Approach

| Design Element | Aim | Solution Approach |
|---|---|---|
| Context | -How an SRA is defined? -Where SRA can be used? | -SecDOAR SRA is defined for standardised architectural representation. -SecDOAR is defined by leveraging existing literature and industrial tools. |
| Goal | -What is goal of an SRA? | The SecDOAR provide a high-level security data architecture along with security metrics which can used for capturing, analysing and reporting security data. |
| Maturity State | -What is maturity stage of a SRA? | The SecDOAR SRA is a standardisation attempt to fill the gap in existing architecture and commercial tools. |
| Design Strategy | -What is described in a SRA? -How an SRA is described? -How elements of an SRA are presented? | -The SecDOAR SRA is presented in terms of components and relation among the components. -The SecDOAR SRA is described using 4+1 views. -The logical view and process view are represented in terms of meta-models, ontologies and architecture design diagrams. -The deployment view is represented in terms of deployment diagram of the prototype implementation of the SecDOAR SRA. -Composition algorithm for components (or corresponding tools) of the SecDOAR SRA elements for security orchestration, analysis and reporting process. |
| Evaluation | How an SRA is evaluated? | -A comparison of the proposed SecDOAR SRA is performed with existing tools to evaluate the completeness of the proposed SRA. -We have also evaluated the proposed SRA for functional and quality requirements. -We have implemented a prototype of the proposed SecDOAR SRA to demonstrate its feasibility by integrating data from four widely used security tools including Snort [44], Splunk [40], Zeek [42] and LimaCharlie [38]. |
| Instantiation | How an SRA is instantiated? | -We have demonstrated the instantiation of SecDOAR SRA by implementing a prototype system for denial-of-service (DoS) and distributed denial-of-service (DDoS) security attacks data. -The details on implementation can guide readers on how to transform the reported SecDOAR SRA into concrete architectures with respect to specific security orchestration, analysis and reporting requirements of an organisation. |

The financial domain is particularly vulnerable to security loophole exploitation. Therefore, a number of standards have been proposed specifically focusing on the financial domain. Some of these standards are ISO 20022 [4], Society for Worldwide Interbank Financial Telecommunications (SWIFT), Financial information exchange protocol (FIX) and FIX Modelling Language (FIXML). SWIFT security program indicates that developers of a system have to comply with sixteen mandatory security control including eleven software architecture security controls [49]. The security requirements include strong access, privileged control, passwords and database controls, multi-factor authentication, control over dataflow links for business processes, effective and robust situational awareness, vulnerability and penetration testing, detection, anomaly analysis and response to security incidents, thorough logging, users activity monitoring and process to audit development of a system. Manual error correction is one of the reasons for introducing potential security vulnerabilities in a SWIFT system because of the highly predictable format of the transactions. A study estimated that fifteen per cent (approximately one billion) SWIFT transactions are manually corrected. Security loopholes in the communication infrastructure of institutions using a SWIFT system is also a potential place for security-related attacks as the SWIFT system is being used in more than two hundred countries by more than eleven thousand financial institutions. FIX and FIXML are used in trading applications [50]. FIXML contains the following elements to describe a financial transaction: (i) details of transaction order, (ii) detail of transaction header and (iii) details of instruments on which transaction is performed and quantity of order. These standards are also prone to the same types of attacks as attacks on the SWIFT systems.

Mandatory and advisory security controls and recommendations have been proposed to counter security threats. The controls for SWIFT infrastructure include [49]: segregation of SWIFT environment from other IT infrastructure of a financial organisation, privileged access control to operating system hosting swift components, internal data flow security of an organisation, reducing the cyber attack surface of SWIFT-related component by providing physical security to control physical access to the systems, password security to have passwords that are resistant to common password attacks, multi-factor authentication, malware protection, software integrity, database integrity, logging and monitoring, cyber incident response planning, and security training and awareness.

Figure 2 summarises different data security standards and corresponding data security recommendations proposed by the standards. The security standards are represented as specialisations of the parent security standard or control. Specific element of the corresponding security standards are shown as aggregated elements.

## 5. Proposed Security Meta-models

The meta-models described in this section cover three key aspects of security data orchestration. First, how different concepts associated with security data orchestration are related to each other. Second, how security data is related to security events and how fine-grained descriptions of the security events can facilitate security data orchestration. Third, how security data can be managed by a data processing pipeline to facilitate the security data orchestration among multiple tools.

---
[4]https://www.iso20022.org/

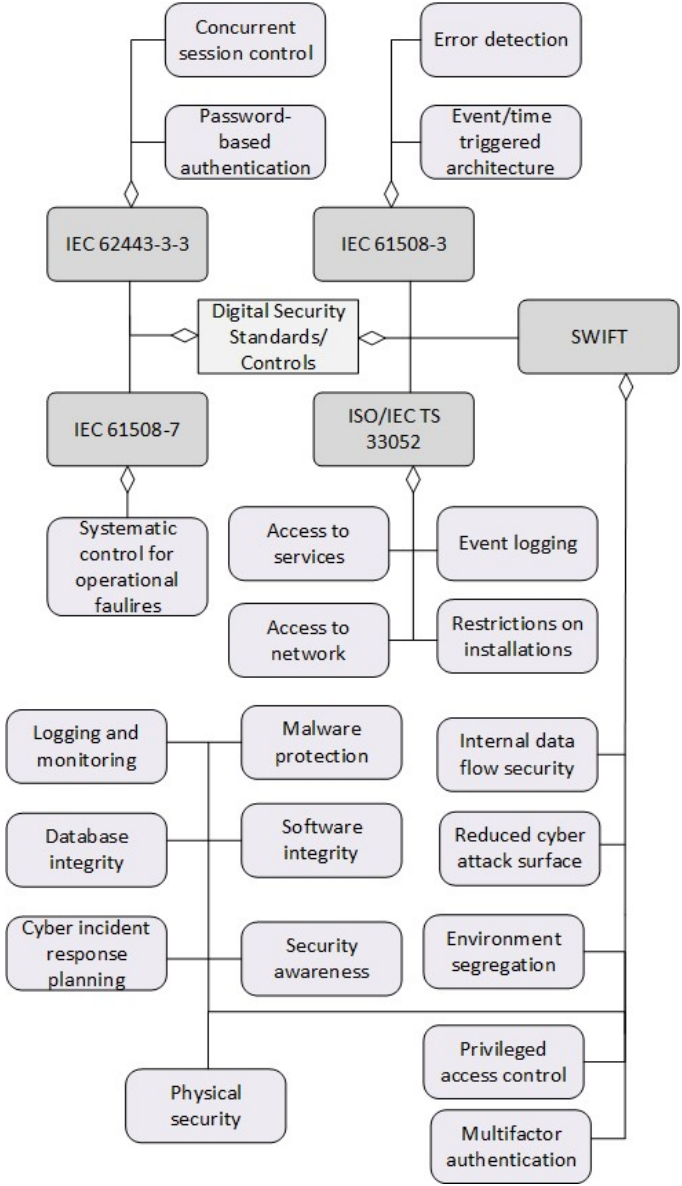

Figure 2: Security Standards and Corresponding Key Security Recommendations

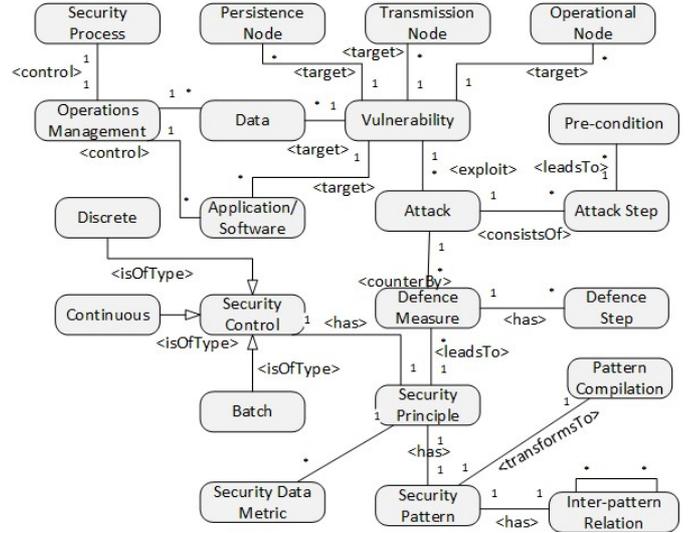

Figure 3: Security Data Meta-model representing Concepts associated with Security Data and corresponding Relations

### 5.1. Security Data Meta-model

A security data meta-model defines relations among different data elements of a security domain [51][52]. A few security meta-models have been reported in the literature to capture security elements of a specific domain. For example, for manufacturing systems dealing with sensitive data, the semantic integration data model encompasses data nodes, operational nodes and transmission nodes [51]. Knowledge sharing among the participating systems is important to support compliance between systems and support handling of heterogeneous data elements [52]. A security mapping data model describes the relations between attack, attack technology, attack step, countermeasures, security principle, security patterns and inter-patters relation. Information structured through a meta-model can be used for generating defence strategies, e.g., for generating attack defence trees [53].

Figure 3 shows a security data meta-model to support Security Data Orchestration, Analysis and Reporting (SecDOAR). The meta-model shown in Figure 3 is an aggregated representation and extension of the aforementioned models and concepts. If a specific security loophole is present in the system and corresponding *pre-conditions* exist, specific *attack steps* can be taken to initiate an *attack* for exploiting a security *vulnerability* in an *application/software* or for accessing corresponding *data*. Security *vulnerabilities* can be exploited in *persistence nodes*, *transmission nodes* or *operational nodes*. Specific *security processes* and *operational management* policies can exist in an organisation that controls the security of the *data* and *applications*. A number of *defence measures* can be taken to counter an *attack* by following specific *defence steps*. The *defence measures* can correspond to high-level *security principles* or concrete security features. A *security principle* or a security feature can have a number of associative *security patterns* or tactics. A *security principle* can also have associated metrics that can be used to capture the data for the related security attacks and perform security analysis using the captured data. The *security patterns* are related to each other via *inter-pattern relations*. The *security patterns* can be adapted or transformed into specific implementations by pattern compilation. Each security pattern can have an associative *security control*, which determines whether the application of a specific security control is *discrete*, *continuous* or to be executed periodically as a *batch* process.

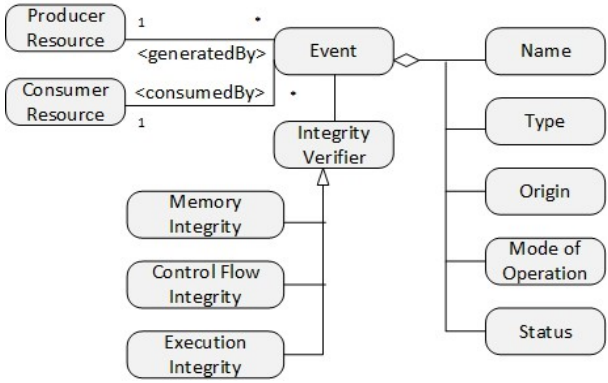

Figure 4: Security Events Meta-model representing Attributes of a Security Event and Purpose/Usage

*5.2. Security Event Meta-model*

Incorporating an event-driven architecture in security analysis and reporting platforms can facilitate timely notification of security incidents and triggering corresponding security responses. The first step towards incorporating an event-driven architecture for data security is to identify attributes that are associated with different events generated by a system under observation [54][55]. These attributes can include but are not limited to: (i) name of the security event, (ii) type of the event, (iii) location of the event, (iv) resources associated with an event, (v) status of the event such as mode of operations, current tools or components associated with the events, and (vi) consumption of the resources corresponding to the cause of the event (e.g., energy, duration etc.) [55]. Endpoints can be attached to different data sources or applications dealing with data. Endpoints can include the data adapters, which can transform one type of data into another. Endpoints can also be used to filter events before these are broadcasted on the enterprise service bus.

Event-driven monitoring can be performed when specific events occur and can be used for maintaining a log of the security incidents [56]. Data hooks are to be placed within data processing components, data storage components such as storage tables or within control flows. The integrity of the event-driven monitors can be managed in different ways [56]:

- By verifying memory integrity to make sure that memory is not corrupted.
- By verifying control flow integrity to make sure that control flow is not maliciously altered.
- By verifying execution should be monitored through a designated entry point.

In addition, the interceptions of instruction execution, interruption of tasks and exception handling flows is also important.

Figure 4 shows the event meta-model corresponding to the SecDOAR SRA presented in this paper. As shown in the Figure 4, an event can be comprised of different attributes including name, type, origin, model of operation and status. An event can be associated with a producer, or one or more consumers of a specific data resource. An event can be associated with multiple integrity verifiers for verification of memory, control flow and execution integrity.

*5.3. Security Data Management Process Meta-model*

Some software security standards provide insights to security data orchestration, analysis and reporting processes. For example, ISO/IEC 27034:2011[5] standard identifies that security is context dependent, therefore, a software security process to deal with the security of the ICT infrastructure should cover business, regulatory and technological context. Each business domain can have its specific security challenges, which illustrates the significance of having a business context. A geographical location at which software services are offered can have additional security risks such as restrictions on the use of specific cryptography algorithms or compliance with certain privacy laws, which shows the importance of having a regulatory context. The technological context deals with technology-specific risks, which can arise from using third party software systems or libraries, not having sufficient penetration testing of a system, improper code reviews with respect to the security standards, and unsecured hosting environment of the applications.

Figure 5 shows a pictorial representation of the process for data security understanding and management. Three stages of security-related situational awareness, i.e., *understanding*, *comprehension* and *perception*, along with data associated with the security operations of an organisation contribute towards establishing a *common operating picture* of the security state of the ICT infrastructure within an organisation. Security comprehension stage can aggregate interpretation, heuristics, learn-ing, modelling and simulation for comprehension of the security incidents. Comprehension, perception and common operating picture can have different security data *metrics*. Each security metric has three aspects including metrics elements (or purpose), measurement scheme and response (or reporting). Metric elements facilitate comprehension, measurement scheme facilitates perception, and response or reporting facilitate building a common operating picture. The metrics can be specialised for capturing different types of security data including but not limited to access control, confidentiality, integrity and audits etc. The metrics have been discussed in detail in Section 6.

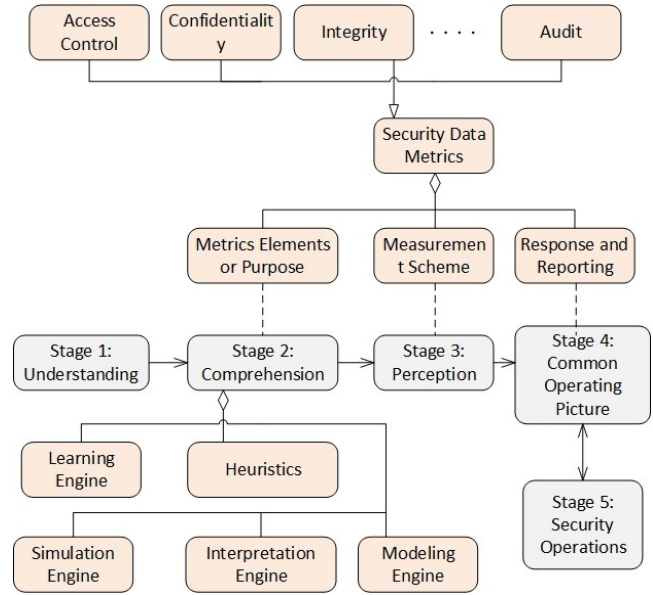

Figure 5: Security Data Management Process Meta-model, Metrics and Interpretation Methods associated with relevant stages of the Process

---

[5] www.iso.org/obp/ui/#iso:std:iso-iec:27034:-1:ed-1:v1:en

## 6. Security Data Intelligence, Information Capturing and Sharing

Security metrics provide a mechanism for collecting data for security analysis and for accessing results of specific security analysis [57]. A catalogue of security metrics for each stage of the software development life cycle can facilitate security by design for logging and monitoring of system behaviour during cybersecurity breaches, and planning for mitigating the security threats in future. A security metric can be defined as a metric whose attributes can be used to measure a security property, indicate if a security has been violated, provide correspondence between a security property (characteristic) and system expected behaviour, and log or record steps that lead to the deviation from an expected system behaviour (in case of security breaches).

For automating the process of security analysis and reporting, security metrics need to be part of the SecDOAR SRA. Incorporating security metrics in an SRA architecture gives a number of advantages [58]. First and foremost, security metrics facilitate the move from the protection of ICT infrastructure towards achieving analysis and reporting capabilities. Secondly, the security metrics identify properties and capture data corresponding to the security threats, which can reduce the number of false positives and help to invoke and apply mitigation strategies on the fly. Thirdly, security metrics can facilitate adding additional sources of data and provide a historical view to analyse security attack patterns over time. The metrics can be used for security incidents and events management (SIEM) as well as for performance and availability monitoring (PAM) [58].

Measuring cybersecurity incidents and developing corresponding metrics can be accompanied by a number of challenges, especially in cyber-physical systems [59]. Selection of the security metrics can be affected by the environment in which a system operates, abstraction level of the collected data that is to be used by the metrics and context in which the analysis results are to be reported (e.g., which tools and technologies are going to be used for visualisation of the results). In addition, knowledge of the system vulnerabilities and available safeguards can facilitate adaptation of the metrics so that the collected data and analysis results can be used for the security threats countermeasures as well. As security can be implemented at multiple abstraction levels in a system and can have a different function at each abstraction level, the security metrics need to be relevant to the abstraction level at which the security is to be monitored. Collecting the data in the relevant security metrics can help to develop a threat landscape, which can be developed incrementally and continuously as adversaries and attackers adapt for new security attacks. Collecting security data using appropriate metrics can also help to quantify losses that occur as a result of a security breach.

Cyber security metrics can be classified into four main categories: physical, informative, cognitive and social [59]. Each of these categories have associative characteristics for preparing plans to respond to a security incident, actions to be taken in a response strategy, strategy to recover from an attack and adapt the system according to the new security configurations [59]. For example, to mitigate physical attacks, probes can be installed to collected physical security related data from critical sensors and system services. An optimised system redundancy strategy can play a significant role in mitigating a security attack. Data about repairing of malfunction controls and sensors and configuration of services in response to past security events also needs to be maintained. In addition, metrics for information security should be able to capture monitoring data from storage as well as access to the services. Effective data transmission from sources to destinations, and review and comparisons of the system data and services before and after the security incidents is also critical. Cognitive security metrics should be able to capture and facilitate analysis of organisational security goals with respect to capabilities of the system to handle security threats. As a result, the metrics can be used to identify critical system services, establish decision making process for recovery options and provide support for reviewing security responses and decision making process. Social security metrics should focus on capturing maturity of the cyber aware security culture, identification of experts to be contacted in case of security incidents, determine liability of the incidents along the organisational hierarchy and evaluate responses of the stakeholders to security incidents to evaluate organisational readiness and communication effectiveness.

For organisations dealing with mission critical and safety critical ICT infrastructure, there can be a need to have an organisation wide process for development and evaluation of the security metrics [59]. An iterative process can be used for setting organisation wide as well as systems specific security objectives, developing metrics, evaluating metrics, combining results from different metrics and evaluating effectiveness of the aggregated results.

Table 2 shows a listing of selected security analysis metrics that are relevant for SecDOAR operations. We have synthesised the metrics discussed in the literature [58][59] in terms of elements, measurement schemes, responses and analysis reporting strategies. These metrics can be used to analyse the aggregated data. The results of the analysis are fed to the reporting/visualisation tool.

*Elements* (or Purpose) of security metrics describe what types of security incidents specific security metrics can capture. For example, metrics for access control can be used to detect incidents of unauthorised access to the system, intrusion detection and breaches of safeguards implemented to incorporate security concerns associated with multi-tenancy (e.g., one

Table 2: Synthesis of Security Metrics, Elements and Adaptation Strategies for Capturing, Analysing and Reporting Security Data and Analysis Results

| Security Metric Focus | Elements or Purpose | Measurement Scheme or Data Collection Strategy | Responses and Reporting |
|---|---|---|---|
| Denial of service and distributed denial of service attacks | -Attack count | -How may times a system has been attacked?<br>-What are sources of attacks?<br>-What ports and hosted systems are under attack? | -Reporting frequency of attacks<br>-Report statistics of network traffic |
| Access control breaches | -Authentication<br>-Non-repudiation<br>-Authorisation<br>-Intrusion Detection<br>-Multi-tenancy or compartmentalisation | -User authentication scheme<br>-User identification scheme<br>-Password implementation or strategy<br>-Successful or unsuccessful password attempts<br>-Monitoring system use<br>-Count of unauthorised access attempts<br>-Login attempts count | -Block unauthorised access<br>-Logging details of targeted systems IPs, components and APIs<br>-Log credentials used for attack |
| Confidentiality and privacy | -Required behaviour<br>-Side channel vulnerability factor<br>-Information leakage measurement<br>-Correlation between attack execution and attack observation | -Maintain system logs | -Tag data that has been targeted in privacy attack<br>-Log details of compromised data and corresponding data owners<br>-Prepare notifications and strategies for reporting privacy breaches |
| Integrity | -Data integrity importance<br>-Integrity Impact | -Calculate violation checks per input or data access request<br>-Measuring impact of vulnerability on a system integrity<br>-Measuring ratio of risky classes or functions with respect to total access classes | -Log integrity violations<br>-Report types and magnitude of integrity violation<br>-Periodic design-time and run-time integrity checks |
| Audit | -Audit trail comprehensiveness | -Track access to data | -Log critical events associated with features and data access as well as data updates |
| Availability | -System services availability | -Denial of service mitigation plan | -Log and report system downtime |
| Source code | -Classified attributes inheritance<br>-Critical class extensibility<br>-Variable vulnerability | -Tracking use of classified attributes<br>-Making critical classes non-extendable<br>-Evaluating security relevancy of variables | -Periodic source code security checks on source code and reporting |
| Version control | -Source code or data changes count | -Tracking how often a source code or data is changed | -Logging and reporting of changes according to the defined rules |
| Data-flow | -Dependency graph | -Tracking how data flows between different elements of the system | -Tagging security and privacy sensitive data, tracking the sensitive data movements and reporting the data movements according to the defined rules |
| Time | -Meantime to fix bugs<br>-Meantime to repair<br>-Attach execution time | -Meantime between discovery and fixing software bugs<br>-Length of time for which an attack has been executed | -Reporting system downtime and losses in terms of data, business and monitory elements |
| Attackability | -Attack count<br>-Attack prone<br>-Vulnerability Index | -Probability of a system to be attacked<br>-Probability of being exploited as a result of an attack<br>-Ratio of fixed vulnerability as compared to discovered vulnerabilities | -Reporting attack surface metrics |
| Software attack surface | -Sensitivity sink<br>-Attack graph probability<br>-Exploitability Risk | -Probability that software or a program can be exploited for security breach<br>-Probability of an attack to succeed<br>-Access to vulnerable code through system interfaces<br>-Probability of sensitive data being exposed as a result of an attack | -Executing and reporting periodic software vulnerability check reports |
| Security feature/requirement incorporated or not | -Incorporated versus total security features/requirements | -Ratio of implemented versus total required security requirements or features | -Logging and reporting system security completion status |

user having access to another user's data).

A *measurement scheme or data collection strategy* describes what type of data should be captured to monitor the activities of users to determine whether a breach has occurred or not. For example, to monitor security characteristics associated with authentication, authorisation, intrusion detection and multi-tenancy compliance, details of every user and actions a user performs should be logged every time a user does something using the system. In addition, the data a specific user has accessed and what modifications have been made in the data should also be monitored.

***Responses and reporting*** deal with what responses can be generated if a security-related anomaly has been detected as a result of analysis after using a specific metric as well as during a recovery strategy from a security breach including system adaptation and system configuration techniques. For example, in case of unauthorised access to the system or a breach of multi-tenancy, the user credentials can be temporarily blocked and information about the user including the IP address through which the request was originated and the piece of data accessed or modified can be logged. These details can be used for further analysis. Table 2 provides details on other metrics that can be used for SecDOAR operations in a similar manner.

## 7. SecDOAR SRA Conceptual Model for Consolidated Representation of the Security Data

The SIEM systems and tools that are used for managing security data and events consist of at least five different features [60]. First, these systems have the ability to monitor target systems that are under observation, and update or reconfigure the target systems to prevent security attacks. Second, these systems have some persistence mechanism that is used for collecting, organising and retrieving security-related data. Third, these systems incorporate mechanisms that support compliance with regulations for the ICT infrastructure in an organisation. This includes support for auditing of the security-sensitive data and associated operations as well as verification and validation of the audits. Fourth, these systems have the capability to relate security events with each other as well as relate security events with the associated security data. This includes the fusion and analysis of the security data with respect to the security events. Fifth, these systems incorporate some type of mechanism for associating security threats with corresponding countermeasures to support incident response planning and related decision making.

A semantic model that captures different aspects of security data orchestration, analysis and reporting needs to have hierarchical representation of information and corresponding operations [60]. Operations on the security data can be classified into multiple categories including detection, correlation, analysis, decision, and reaction or response. Security data can be used for analysis and reporting of security attacks that can target network protocol configurations, system services, operating systems, security signatures, security policies, and security event management. Security event management can include security incidents and attacks logs as well as corresponding security alerts.

Table 3 shows the input and output of the SecDOAR operations. The security orchestration services take security data and integration models (e.g., semantic integration models of security data) as input and produce integrated security data as output. The security analysis services take integrated security data and security analysis metrics as input and produce security analysis results as output. The security reporting services take security analysis results and presentation details as input

Table 3: Input, Output and Objectives of SecDOAR SRA Operations

| Operation | Input | Output | Objective |
|---|---|---|---|
| Orchestration | -Security data -Semantic integration model | Integrated security data | Have a consolidated representation of the security data generated by different security monitoring tools |
| Analysis | -Integrated security data -Security metrics | -Security threat report | Security analysis report indicating presence or non presence of potential security threats |
| Reporting | -Security analysis results | -Results presentation -Reports and reports presentation format | Providing results to analysis tools for further interpretation and actions |

and produce security analysis reports, visualisation and recommended actions as output.

A composition process that can combine security data from multiple input sources, analyse the data and export the result for analysis to multiple output sources is a critical part of SecDOAR process as the security data can be gathered, orchestrated and analysed by a number of tools before actionable information can be extracted from the data. To facilitate the process of security data collection, semantic integration and analysis, we have described the process of SecDOAR using the software process description language proposed in [61].

SecDOAR SRA has a number of elements to support operations such as gathering security data from different tools, analysing the security data using the security metrics and reporting the results of the security analysis to the external tools. Different tools can have different features and the tools can produce and consume data in different formats. A semantic model is required to describe how security data is related to different tools, how security data with the same semantic and different syntax are related, and which specific analysis can be performed on the security data using specific metrics. The semantic model serves as a baseline for security data integration.

Figure 6 shows a semantic model describing a pictorial representation of the SecDOAR SRA concepts and relations among the concepts. Figure 6 shows that security data can be gathered from a number of sources including log files, APIs of the security monitoring tools, security logging information stored in the database, data on runtime behaviour of the different systems, and data that is gathered during system configuration for handling cyber security. *Orchestration Environment* uses security data and represents the data in a manner so that the data gathered from different tools is semantically integrated and the integrated data can be used for further analysis. *Orchestration Environment* can include a number of *Data Channels* to perform data buffering and manage data status. The data can be used by *Security Analysis* components in combination with *Security Tools* for security data analysis and reporting, and *Security Events* for identifying and reporting security breaches. *Se-*

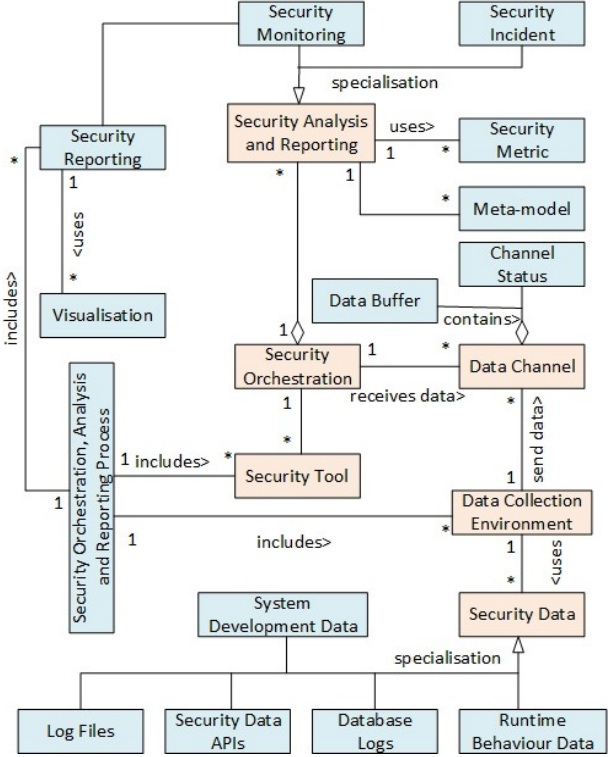

Figure 6: Semantic Composition Model of the SecDOAR SRA

Table 4: SecDOAR SRA Concepts and Corresponding Axioms in Descriptive Logic for System Composition

| SecDOAR RA Concepts | Axioms |
|---|---|
| A tool $C$ has a feature $F_i$. | $\forall C \sqsubseteq \exists Feature.F_i$ |
| A component $C$ has a subcomponent $C_i$. | $\forall C \sqsubseteq \exists subComponent.C_i$ |
| A component $C$ has an interface $I$ can consume data $D$ as an input or output. | $\forall C \sqsubseteq \exists I.D$ |
| A relation $R$ is used to connect two interfaces (or APIs) $I$ and $\bar{I}$. | $\forall I \sqsubseteq \exists R.I$ |
| A component $C$ is associated with another component $C$ via an interface $I$. | $\forall C \sqsubseteq \exists I.C$ |
| A security data $D$ has a data collection environment, which is associated with a data channel that can contain data buffer channel and state $S_D$ of data buffer channels $S_a$. | $\forall D \sqsubseteq \exists DataCollectionEnvironment_D \sqsubseteq \exists DataChannel_D, \sqsubseteq \exists DataBuffer_D, \sqsubseteq \exists S_D$ |
| A security data $D$ can be capture via security log files ($D_{Log}$), security data APIs ($D_{API}$), database logs ($D_{DatabaseLogs}$), applications runtime behaviour ($D_{RunTime}$) or system development data ($D_{Development}$). | $\forall D \in \{D_{Log} \lor D_{API} \lor D_{DatabaseLogs} \lor D_{RunTime} \lor D_{Development}\}$ |
| Security data analysis and reporting process $Process_{AR}$ uses security metrics $Security_{Metrics}$ and meta-models $Security_{Meta-models}$. | $Process_{AR}(Security_{Metrics}, Security_{Meta-models})$ |
| A security data $D$ has an active state $S_a$. | $\forall D \sqsubseteq \exists hasState S_a$ |
| A security data $D$ has an inactive state $S_i$. | $\forall D \sqsubseteq \exists hasState S_i$ |
| Two security data $D_x$ and $D_y$ are related to each other. | $\forall D_x \sqsubseteq D_y$ |
| Two security data $D_x$ and $D_y$ are not related to each other. | $\forall D_x \sqsubseteq \neg D_y$ |
| Two types of security data $D_x$ and $D_y$ can be orchestrated using a semantic integration model $SIM_i$ using a function $F$, which results in an integrated data $D_z$. | $D_z \equiv F_O(D_x, D_y, SIM_i)$ |
| A security analysis metric $M_i$ can be applied on security data $D_z$ using a function $F$ to produce analysis results $AR_i$. | $AR_i \equiv F_A(M_i, D_z)$ |
| A security reporting method $R_i$ can be used to report an analysis result $AR_i$ using a function $F$ to present results $Res_i$. | $Res_i \equiv F_R(R_i, AR_i)$ |
| A component $C$ description in terms of its properties and relation to data. | $\forall C \equiv \exists subComponent.C_i \sqcap \exists I.C \sqcap \exists R.\bar{I}\exists I.D_x \Rightarrow D_y \sqcap (\exists hasState S_a \lor \exists hasState S_i)$ |
| A security tool $T_x$ containing a component $C_x$ providing a security feature $F_x$ to support an orchestration $F_{Ox}$, analysis $F_{Ax}$ or reporting $F_{Rx}$ function. | $\forall T_x \sqsubseteq \exists component.C_x \sqsubseteq \exists feature.F_x \exists supportFunction.(F_{Ox} \lor F_{Ax} \lor F_{Rx}) \exists hasInterface.I_x \exists consumesData \underline{D_x} \exists producesData D_x)$ |
| A composition description to provide a specific user case for security orchestration, analysis and reporting using tools $T_i$, $T_j$ and $T_k$ respectively in terms of components involved, data exchanged and features provided. | $\exists composition.(T_i \land T_j \land T_k) \sqsubseteq ((T_i \sqsubseteq \exists component.C_i \sqsubseteq \exists feature.F_i \exists support.F_{0i} \exists hasInterface.I_i \exists consumesData \underline{D_j} \exists producesData D_i) \Rightarrow (T_j \sqsubseteq \exists component.C_j \sqsubseteq \exists feature.F_j \exists support.F_{Aj} \exists hasInterface.I_j \exists consumesData \underline{D_j} \exists producesData D_j) \Rightarrow (T_k \sqsubseteq \exists component.C_k \sqsubseteq \exists feature.F_k \exists support.F_{Rk} \exists hasInterface.I_k \exists consumesData \underline{D_k} \exists producesData D_k)) \land (D_i \sqsubseteq D_j \sqsubseteq D_k)$ |

*curity Events* reporting takes input from one or more *Security Metrics*. *Security Events* can have three specialisations for handling details of the *Security Incidents*, algorithms for *Security Monitoring* using the security data, and *Security Reporting* to report if a security breach has been identified. *Security Reporting* can be done using a number of visualisation techniques and security reporting tools for presenting details of the identified or potential security breaches. Axioms presented in Table 4 can be used to evaluate different instances of the SecDOAR SRA by using the semantic model to find an optimal composition of the components for a specific security monitoring scenario.

Describing SecDOAR SRA elements using descriptive language axioms provides a formal structure of the concepts of the SecDOAR SRA and a mechanism for identifying SecDOAR system composition requirements and verifying whether a specific instance of the SecDOAR SRA meets the desired SecDOAR requirements. Table 4 describes relationships between concepts of the SecDOAR SRA and corresponding axioms. The descriptive logic axioms can be used to describe the structure of the SecDOAR SRA elements and their composition in a specific instance of the SecDOAR SRA. Table 4 also describes elements of SecDOAR semantic composition meta-model shown in Figure 6. The key concepts discuss in Table 4 are tools, components, security data and security data analysis metrics. Tools are at the highest level of abstraction for the SecDOAR process. Different tools can be provided by different vendors and support different security data analysis features. Each tool can consist of several components or services through which it can share security data with other tools or con-sume security data produced by other tools. The security data can be analysed by the tools using different security analysis metrics.

## 8. SecDOAR SRA Components and Elements

The high-level architecture for Security Data Orchestration, Analysis and Reporting (SecDOAR) encompasses several ca-

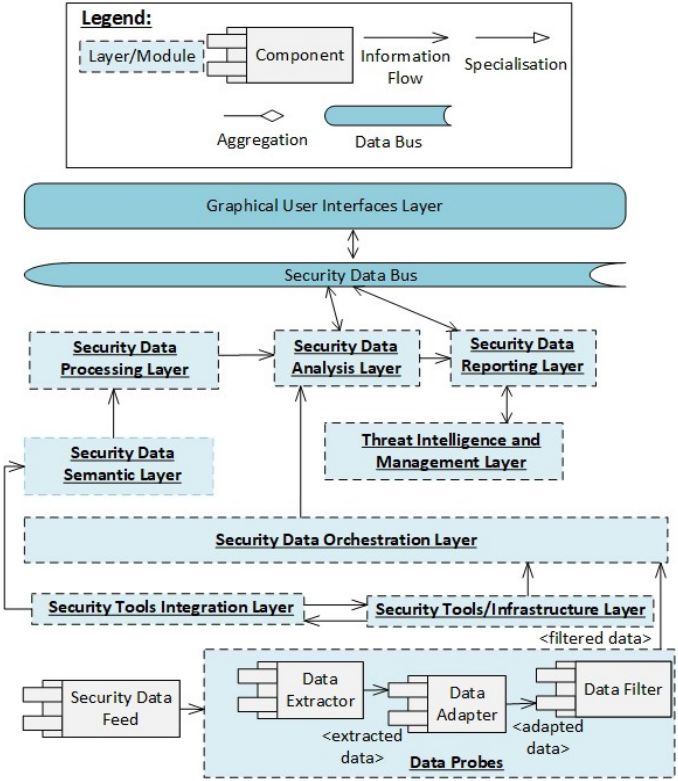

Figure 7: SecDOAR SRA High-level Architecture showing Layers, Components and Information Flow

pabilities related to security data representation, semantic integration, security data extraction, cleaning, loading, analysis, reporting and combining event-driven architecture with analysis and processing components of the architecture. The SecDOAR SRA reported in this paper has been through the integration of existing architectures to provide standardisation, therefore we have described the synthesis of SecDOAR SRA. The synthesised SecDOAR SRA provides an aggregated representation of all the concepts discussed in this paper. Figure 7 shows a high-level representation of SecDOAR SRA elements and components. The legend shown in Figure 7 is used for the details of layers shown in the diagrams of the corresponding layers of the SecDOAR SRA.

### 8.1. Security Tools and Infrastructure Layer

The Security Infrastructure Layer manages capabilities and activities supported by different security tools. Three types of security tools are used for security operations. *Security anomaly detection* tools are used to monitor an organisation's ICT infrastructure for capturing data related to potential security breaches. *Security analysis* tools are used to analyse the captured data. *Security reporting* tools are used to present the results of the analysis to the security operation team or external tools. There can be one or more security tools in a SIEM platform. Each *security tool* can have different capabilities (e.g., which types of security data a tool can capture) and support different types of security activities (e.g., what type of security operations a tool can support, such as a firewall can block specific ports if the ports are being targeted by an attack). Security *management process* combines one or more tools in a pipeline so that security data among the tools can be exchanged. Figure 8 shows the components of the security tools and infrastructure layer.

### 8.2. Security Integration Layer

The Security Tools and Integration Layer handles integration among the security data originating from heterogeneous sources and security data analysis tools by providing *Application Programmable Interfaces (APIs)*, *plug-ins* and *tools registry*. The details of the tools integration layer are shown in Figure 9(c). The tools registry is used to register the tools with the SecDOAR platform. Plug-ins and APIs are used to feed the data by the security tools to the SecDOAR platform.

### 8.3. Semantic Layer

Security data from different tools is fed to the *Semantic Layer* for semantic integration of the data captured by different tools. The details of the components in this layer are shown in Figure 9(a). The *Ontologies or Knowledge Base* stores ontologies to guide semantic representation of the security data. The *Query Engine* facilitates the extraction of semantically related security details from the security data using ontologies and knowledge representation. The *Information Extractor and Interpreter* extracts reusable information or knowledge.

### 8.4. Security Data Processing Layer

The *Security Data Processing Layer* extracts additional information about security incidents from the raw data provided by the security tools and persists the data so that it can be used for analysis at a later stage. This layer consists of three components as shown in Figure 9(b). The *Data Curator* component saves the captured raw data. The *Data Extractor* component extracts additional information about the security incidents from the raw data. The *Data Persistence* component saves the extracted information for later use.

### 8.5. Threat Intelligence and Management Layer

The *Threat Intelligence and Management* layer uses security data as input and analyse the data for identification of different types of threats and provides analysis using the *Security Data*

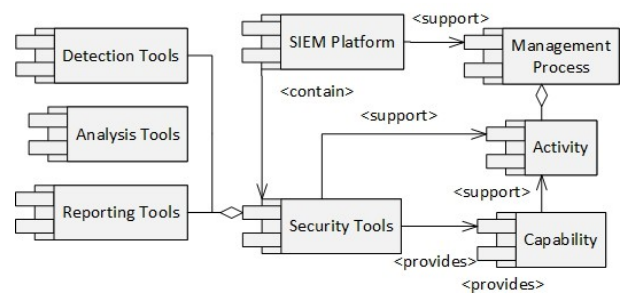

Figure 8: Security Tools and Infrastructure Layer

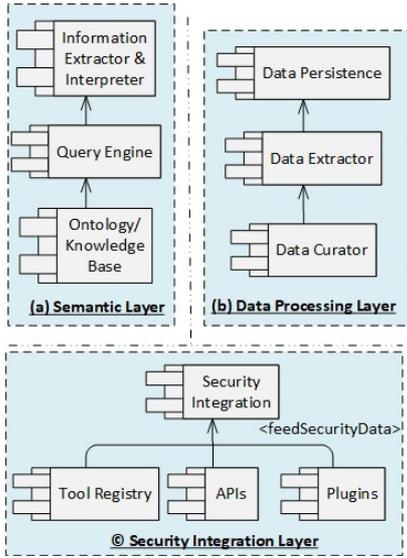

Figure 9: Semantic Integration, Data Processing and Integration Layer

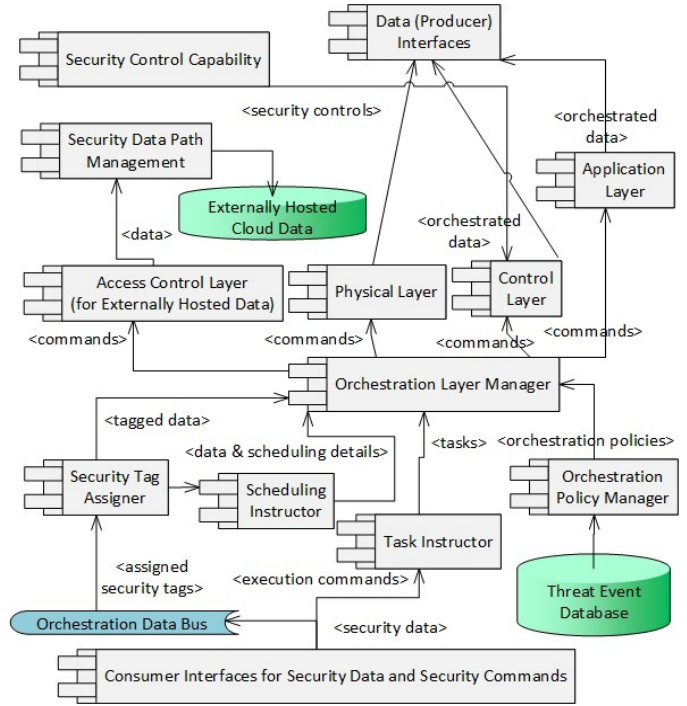

Figure 11: Orchestration Layer Detail

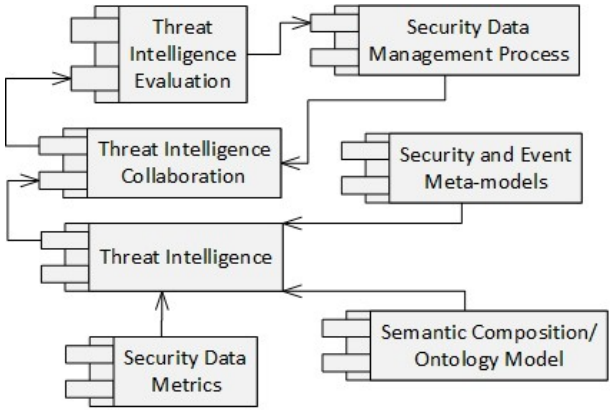

Figure 10: Threat Intelligence and Management Layer

*Metrics* (as discussed in Section6). The *Security and Event Meta-models* and *Semantic Composition Model* facilitate relating different types of security data and events. The gathered security data is passed on to other system entities for *Threat Intelligence Collaboration*. The *Threat Intelligence Evaluation* and *Security Data Management Process* guide the selection of information to be shared for collaboration. Figure 10 shows the details of threat intelligence and management layer components.

*8.6. Orchestration Layer*

The Security Data Orchestration layer handles how security data from multiple tools and SIEM platforms can be orchestrated. The orchestration process is also encompassed in the orchestration layer and leverages organisational security management processes for adapting appropriate abstraction for a security orchestration process. Figure 11 shows details of the orchestration layer. External tools and data sources are integrated with the platform via *Interfaces for Security Data and Commands*. In the next step, the security data are tagged with security information, the tagged data can be used for *Scheduling Instructions*. *Task Instructions* and *Orchestration Policy Manager* are used in conjunction with the scheduling instructions for providing security data orchestration by *Orchestration Layer Manager*. The output of *Orchestration Layer Manager* is used by access control, physical and application layer of the security data analysis and reporting tools. The output from the orchestration layer can be used for two purposes. Firstly, it can be used for storing security data on external storage services. Secondly, it can be used for exposing orchestrated data via interfaces to the *Physical*, *Control*, *Application*, *Access Control*, *Security Data Path* and *Data Interface* components.

*8.7. Analysis Layer*

Figure 12 shows the components of the analysis layer and their interaction. The first part of the analysis layer works as follows. *Security Controls Capability* takes input from *Security*

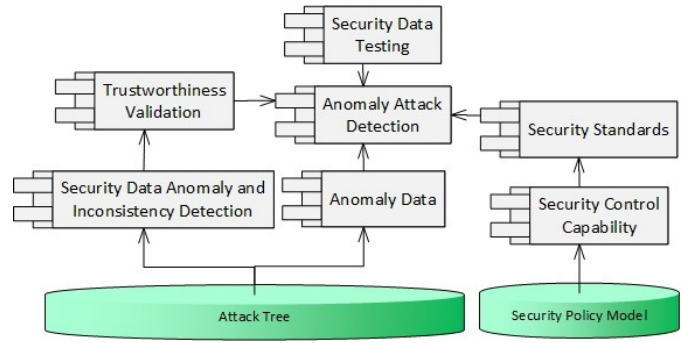

Figure 12: Analysis Layer Detail

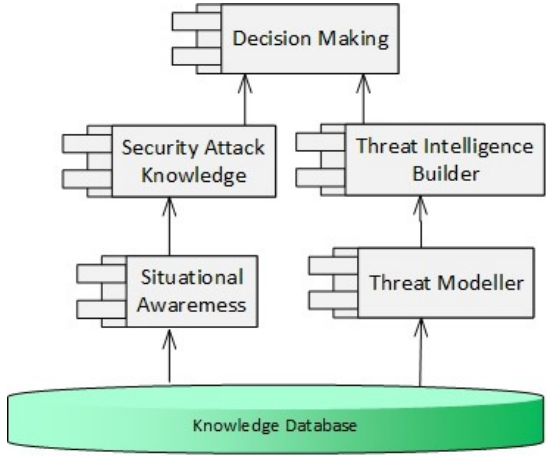

Figure 13: Reporting Layer Detail

*Policy Models* and gives details to *Security Standards*. Attack defence trees can provide details of the attacks to the *Security Data Anomaly and Inconsistency Detection* service. If inconsistencies are detected by these services, then *Trustworthiness Validation* service validates whether the generated results can be trusted. In the next step, the *Anomaly Attack Detection* service prepares the analysis results with the help of the *Security Data Testing* service.

*8.8. Reporting Layer*

Figure 13 shows the components of the security data reporting layer. *Knowledge Database* contains security analysis results. The results can be used for *Threat Modelling* and *Threat Intelligence Building* so that an appropriate reporting mechanism can be adopted. At the same time, *Situational Awareness* of the current security situation and *Security Attack Knowledge* present security metrics that are used and corresponding security threats that are focused on by the metrics. Output from all of these components is used for *Decision Making* for security reporting.

## 9. SecDOAR SRA Evaluation and Instantiation

In this section, we report an evaluation of the presented SecDOAR SRA by demonstrating the feasibility of the SecDOAR SRA with the help of an implementation case study and in terms of SecDOAR SRA support to fill the gap in existing commercial products.

*9.1. Implementation Case Study*

We have instantiated SecDOAR SRA as a prototype platform for proof of concept and evaluation case study using open source or trial versions of Snort [44], Splunk [40], Zeek [42] and LimaCharlie [38] tools. These tools have been chosen because these are widely used to gather security of enterprise systems. These tools are used to monitor network security and gather network security-related data. The purpose of carrying out the case study was twofold. Firstly, to analyse the feasibility of the proposed SecDOAR SRA for real-life security scenarios using existing security monitoring and data collection tools. Secondly, to evaluate the capability of the proposed semantic model and composition approach for (i) orchestrating security data being gathered from heterogeneous sources, (ii) using security metrics to be able to perform some analysis on the data, and (iii) reporting of the results for subsequent actions.

*9.1.1. Evaluation Scenario and Tools Composition for Security Data Capturing*

The scenario used for prototype evaluation focused on detecting attacks that can lead to denial-of-service (DoS) and distributed denial-of-service (DDoS) attacks. We selected targetted tools and their features that can provide DoS and DDoS security data. The security data elements related to DoS and DDoS include the IP address of the target system as well as the system through which the request has originated, the port of the target system at which the request is received, and the size of the response data. We chose three tools Snort [44], Splunk [40] and LimaCharlie [38] to gather security data that can provide information on DoS and DDoS attacks. After applying the composition model described in Table 4, the following composition of the tools is derived.

$\exists composition.(Snort \land Splunk \land LimaCharlie) \sqsubseteq ((Snort \land Splunk \land \underline{LimaCharlie}) \sqsubseteq \exists feature.(DoS \land DDoS) \exists producesData.(DoS, DDoS))$

*9.1.2. Driving Integration Ontology from Semantic Composition Model*

The conceptual integration model presented in Section 7 has been used to build the semantic integration model of the prototype. The graphical representation of the instantiated semantic integration model is presented in Figure 14. The semantic model captures a use case in which data is sent over a network between a source and a destination, monitored by security tools for security threats. *Traffic data* has the following attributes for the use case implemented in the case study: IP address and port of the source, IP address and port of the destination, underlying network protocol, priority for observing the network traffic, the content of the data and temporal elements. *Traffic metric* monitors the *traffic data*, e.g., how many times an anomaly or a threat has been detected. *Traffic visualiser* uses *traffic data* and *traffic metrics* to represent the results for subsequent decision-making.

*9.1.3. Selecting Security Metrics for Analysis*

The scenario for evaluation focuses on DoS and DDoS attacks. The following security data and metrics identified in Table 2 are used in the prototype implementation.

- Frequency of request on a specific IP address and port that listen to an incoming request for a hosted application is used to distinguish normal traffic patterns from abnormal traffic patterns.

- The IP addresses from which the information originated can be used to determine whether the request has originated from a legitimate node or an attacker node.

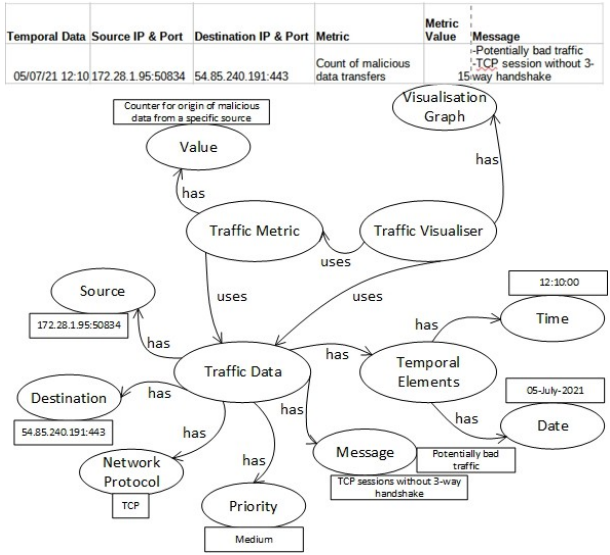

Figure 14: Semantic Model for the Case Study Data and Tools

- User credentials (user name, password and access codes) used to access an application can be used to identify if particular user credentials or a group of credentials are being exploited.

By combing the aforementioned metrics, we can determine whether an HTTP request (or access) to a system is part of DoS or DDoS. For example, a system hosted on a machine with IP address 54.85.240.191 and listening on port number 443. If a normal request pattern for the IP address and port number is less than 15 request per second, a threshold can be defined that if there are more than 15 requests per second, the system might be a target of DoS and DDoS attack. A metric identifying whether the requests originate from same client IP addresses or different IP addresses is another metric that can be used to identify a potential DoS and DDoS attack. Having detailed analysis of the credentials in the attacks can provide an insight whether the users data (credentials, passwords and access codes) have been compromised or not.

In our case study, we have only focused on metrics that are associated with HTTP requests to the system. We do not focus on the contents of the parameters passed to the server as that is outside the scope of the research presented in this paper on SecDOAR SRA and pipeline.

### 9.1.4. Selecting Visualisation for Reporting

We have focused on DoS and DDoS attacks in the prototype implementation of SecDOAR reference architecture. Therefore, we have selected graphical visualisations that can provide details on DoS or DDoS attacks including:

- IP addresses from which potential attacks originated. This visualisation includes an overview of all the distinct IP addresses from which potential attacks originated.

- IP addresses from which failed attempts to access system services were made. This visualisation includes wrong authentication credentials or attempts to access web services via wrong interfaces or by providing wrong data formats to the web services.

- How many successful or failed attempts have been made to access a system's services from an IP address over a specific time period? This visualisation provides a summary of traffic patterns from an IP address to determine whether the behaviour of the client is normal or of a potential attacker.

Using the aforementioned visualisation can provide an insight into potential DoS or DDoS attacks for the scenario of our prototype. For other scenarios, other visualisations might be more suitable.

### 9.1.5. Instantiating SecDOAR SRA for Implementation of the Platform

We have implemented a prototype of the SecDOAR SRA platform as a Service Oriented Architecture (SOA) [62] by selecting some of the key components of the SecDOAR SRA. We have used .NET[6] and JavaEE[7] frameworks including JaxRS[8], JaxWS[9] and ASP.NET Web APIs[10] for implementing the web services. We have used a mixture of JavaEE and .Net frameworks to demonstrate that SecDOAR SRA can be instantiated in a platform-neutral manner. We have used Apache Jena Framework[11] and dotNetRDF[12] for implementing the semantic model. SPARQL[13] query language is used to fetch the data from the Resource Description Framework (RDF) representation of the semantic model. The tools and services of the platform have been hosted on either Virtual Machines (VMs) installed on servers, Azure Virtual Machines [14] instances or in-house Openstack[15] private cloud instances.

Figure 15 shows a deployment view of the instantiated architecture. We have implemented instantiated components in the prototype as web services. In the prototype implementation of the SecDOAR SRA, we have only instantiated components relevant to our use case. The security tools deployed on VMs monitor target ICT infrastructure. The data from the tools are captured using the probes encapsulated by *Data Parser/Transformation Service*. *Security Data Structuring Service* collects the data in tools-specific format, transforms the data into a structure that can be recognised by the target metrics and stores the data in intermediary tables in the database. This service uses security metrics from *Security Metrics Service* to take input on the structure of the data. To keep the data gathering process simple, we have used a separate intermediary table for each tool used in the prototype. *Data Orchestration Service* fetches the data and structures it in a semantically related

---
[6]https://dotnet.microsoft.com/
[7]https://www.oracle.com/au/java/technologies/java-ee-glance.html
[8]https://docs.oracle.com/javaee/6/tutorial/doc/giepu.html
[9]https://docs.oracle.com/javaee/6/tutorial/doc/bnayl.html
[10]https://dotnet.microsoft.com/apps/aspnet/apis
[11]https://jena.apache.org/
[12]https://dotnetrdf.org/
[13]https://www.w3.org/TR/rdf-sparql-query/
[14]https://azure.microsoft.com
[15]https://www.openstack.org/

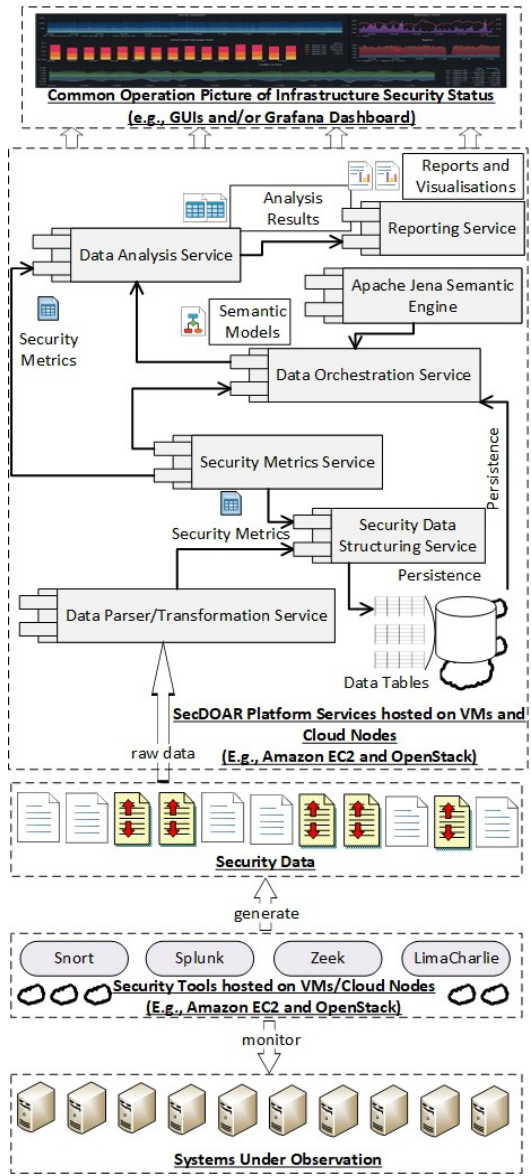

```
 1  [
 2    {
 3      "failedLoginCount": 0,
 4      "portList": "4563;4219;3025;4407;3714;
 5      "startTime": "2021-07-20T00:15:04",
 6      "endTime": "2021-07-20T00:15:04",
 7      "sourceIp": "201.3.120.132",
 8      "destIP": "172-31-27-153",
 9      "attemptNum": 27
10    },
11    {
12      "failedLoginCount": 0,
13      "portList": "2847;3547;2979;2105;3085;
14      "startTime": "2021-07-20T00:15:04",
15      "endTime": "2021-07-20T00:15:04",
16      "sourceIp": "76.169.7.252",
17      "destIP": "172-31-27-153",
18      "attemptNum": 26
19    }
20  ]
```

Figure 16: Aggregated Data for Login Attempts - Example

Figure 15: SecDOAR SRA Platform Instantiation - Platform's Deployment View

knowledge base using the semantic models implemented in *Semantic Engine* using Apache Jena Framework and dotNetRDF APIs. Figure 16 shows an example of aggregated data that can be used for tracking access to the systems (physical or virtual machine as well as hosted applications listening on a specific port). The semantically related data along with security metrics are provided to the *Data Analysis Service*, which performs analysis on the semantically related data using SPARQL queries and computes results on the security status of the observed ICT infrastructure using the security metrics. The analysis results are used by *Reporting Service*, which transforms the results into a presentable format. The results are shown to security operations, configurations and infrastructure management teams using tools such as Grafana dashboards[16] or customised graphical

---
[16]https://grafana.com/grafana/dashboards

user interfaces (GUIs) for visualisation of the results.

Figure 17 shows GUIs used for the visualisation of the security analysis results for DoS and DDoS attacks. The GUIs we have shown in the figure use a threshold of 20 requests per second along with other criteria to classify a network traffic pattern as a potential attack. This threshold can be changed according to the requirements of the scenarios. Figure 17(a) shows a summary of IP addresses that are being targeted in a security attack. Figure 17(b) shows the percentage of the invalid access attempts to the systems which receive more than 20 requests per second. Figure 17(c) shows suspicious IP addresses from which potential security attacks have originated. A cyber security incident response team can use the reporting metrics and associated GUIs to make decisions for enhancing the security of ICT infrastructure.

### 9.2. Evaluation in Terms of Support for Security Operations

We have analysed SecDOAR SRA in terms of its support for security operations performed by security orchestration platforms. We have selected some of the widely used commercial platforms/systems for our analysis including DATADOG [63], SolarWinds Security Event Manager [64], Manager engine event log analyser [65], Splunk [40], OSSEC [66], LogRhythm Platform [67], AlienVault Unified Security Management [68], NetWitness Platform [69], IBM QRadar SIEM [70] and McAfee Enterprise Security Manager [71].

Table 5 summarises how different elements of the SecDOAR SRA can support SOAR operations. We have mapped salient features of each commercially available tool on elements of the SecDOAR SRA to indicate (i) how incorporating implementation of the SecDOAR SRA in the security operations infrastructure can facilitate security data interoperability, and (ii) how SecDOAR SRA presented in this paper helps to fill the gap in

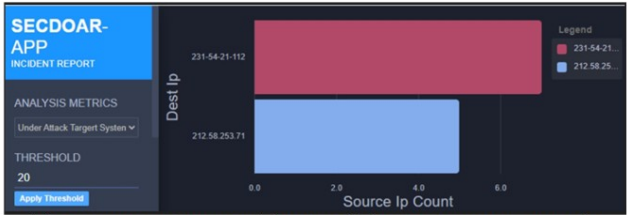
(a) GUI summarising IP addresses of destination systems under potential attacks

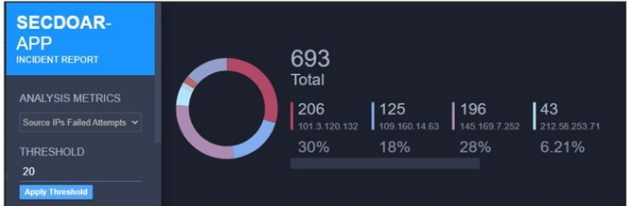
(b) GUI summarising source IP addresses of potential attacks

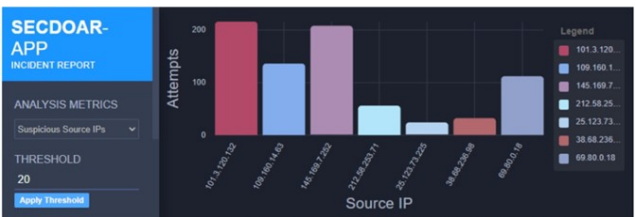
(c) GUI summarising count of attacks originated from specific source IP addresses

Figure 17: Security Data Reporting GUIs

the available commercial solutions. As shown in the table, commercial tools lack support for thorough security data integration, which can be complemented by either incorporating SecDOAR SRA in the tools or having an underlying platform (as demonstrated in an implementation case study in Section 9.1) that can support integration among the tools. By incorporating security data integration support in the security infrastructure, the organisations can integrate multiple tools as per their security needs without allocating additional resources on explicit tools integration, which is often difficult to achieve when tools are provided by different vendors.

## 10. Conclusions and Directions for Future Work

In this paper, we have presented a Software Reference Architecture (SRA) for Security Data Orchestration, Analysis and Reporting (SecDOAR). We have presented meta-models for security data, security events and security data management processes. In addition, we have identified high-level and detailed components of the SecDOAR SRA. Moreover, we have also instantiated the SecDOAR SRA by implementing a prototype of the SecDOAR SRA using security data generated from a selected set of security tools.

The aim of the research presented in this paper has been to provide an architecture-centric solution to support interoperability between heterogeneous security tools based on the security data. While investigating the problem, designing the reference architecture and implementing a prototype, we made a number of observations that can provide insight into the challenges of the security data orchestration domain and respective strategies to address the challenges.

Security data generated from different tools can be associated with security concepts at different levels of abstraction. One of the challenges for providing security data orchestration is *to relate security data concepts at different levels of abstraction in a meaningful manner*. For example, in order to associate the security data gathered from multiple tools with a security vulnerability, a SecDOAR platform needs to identify security controls. The security controls are used to gather the data, extract information about the attacks that are identified using the data, and report vulnerabilities in an organisation's Information Communication and Technology (ICT) infrastructure that are exploited by the attacks. To address this issue, we have proposed a security data meta-model that captures relations among security concepts at different levels of abstraction.

*Security events associated with the security data and the process through which security data is managed* are important for SecDOAR operations. The security data-capturing process can be triggered as a result of certain actions. For example, when there are repeated unsuccessful attempts to access a system from specific IP addresses, or some web service is being targeted by exceptional volumes of input data or requests. Each of such actions can have a corresponding security event. In addition to the identification of the security events, three main elements of the security events are vital. These elements are (i) why a security event was triggered, (ii) at which stage of the SecDOAR system operations a specific security event can be relevant, and (iii) what is the current status (active or expired) of a security event. Moreover, a security event needs to capture the source and target of the security attack as well as how integrity (e.g., control flow integrity or execution integrity) associated with the security data can be maintained. *A comprehensive and well-structured security events model can complement security data management processes.* The security events can be associated with different stages of the SecDOAR processes including but not limited to comprehension, perception and operationalisation.

Security data metrics are a vital part of any SecDOAR platform. For security metrics to be effective for SecDOAR operations, these metrics should have the following elements. (i) *Properties of the metrics* in terms of what *security characteristic* these metrics can capture. For example, access control metrics can be used to validate authentication, non-repudiation, authorisation, intrusion detection and breach of multi-tenant data access. (ii) What kind of *measurements* can be incorporated in a security metric to capture the security data? For example, the access control metric needs to capture successful and unsuccessful password attempts, count of unauthorised access attempts, and any changes in the access control strategy that can result in the breach of authentication integrity. (iii) What are *responses and reporting mechanisms* for analysis of the data and exporting results of the analysis to external tools for deploying security threat mitigation strategies? For example, mitigation strategies to counter system access related-attacks can include temporarily blocking credentials used for unauthorised access, logging details of IP address from which an attack is originated, logging IP address, ports number and details of the APIs targetted during a security attack.

Table 5: SecDOAR SRA Comparison and Correspondence with the selected SIEM Tools

| SIEM Tools | SecDOAR SRA Elements and Components | | | | | |
|---|---|---|---|---|---|---|
| | Data Meta-model | Event Meta-model | Process Meta-model | Metrics | Semantic/ Integration Model | Components |
| DATADOG [63] | ✗ | -Authentication events | -Security compliance control | -Network traffic monitoring -Ports monitoring -Brute force attacks -Critical API calls | Security data from multiple sources | -Visualisation |
| SolarWinds Security Event Manager [64] | ✗ | -Event log | -Security compliance management -Compliance reporting | -Persistence threats -Data log analyser -DDoD attack detection - SQL injection tracking -User activities -Network monitoring | ✗ | -Analysis and reporting |
| Manager engine event log analyser [65] | ✗ | -Event log | ✗ | -User session monitoring -Network device monitoring -Network auditing | ✗ | Historical security event trends analysis |
| Splunk [40] | ✗ | -Security events | ✗ | -Risk identification | ✗ | -Alert management -Alert scores -Security response automation |
| OSSEC [66] | ✗ | ✗ | ✗ | -Network intrusion detection -log file management | ✗ | ✗ |
| LogRhythm Platform [67] | ✗ | ✗ | ✗ | -System usage logs -Data flow logs -Application usage logs -Audit logs | ✗ | -Identify security threats |
| AlienVault Unified Security Management [68] | -Threat exchange | -Security event | -Security regulations compliance | -System usage logs -Log of data generated by systems -Intrusion detection -Data integrity monitoring -Network traffic monitoring -Events logs integrity -User behaviour monitoring | ✗ | Co-relating security data with security vulnerabilities |
| NetWitness Platform [69] | ✗ | ✗ | ✗ | -Logs management -Network traffic detection -Endpoint detection | ✗ | -Security orchestration |
| IBM QRadar SIEM [70] | -Activities and incidents correlation -Threat intelligence and support for STIX/TAXII | ✗ | ✗ | ✗ | ✗ | Detection and prioritisation of threats across organisation |
| McAfee Enterprise Security Manager [71] | -Monitoring and analysing multi-source data including STIXX feeds | ✗ | ✗ | -Activity monitoring -Threat intelligence | ✗ | Heterogeneous data monitoring and intelligence |

**Semantic interoperability of the data** can support aggregated operations on different types of security data. A semantic model used for security orchestration should be able to capture not only relations among different types of data and sources of security data but also how the security data can be used by different security tools for analysis. To achieve this, we have provided *a semiformal approach to express SecDOAR SRA elements in descriptive logic so that the dynamic composition of the security tools based upon the security data needs can be supported*. Descriptive logic to express security data and corresponding tools can provide a verifiable security tools composition framework, which can be tailored according to the specific needs of a security operations team.

Our experience with instantiating the SecDOAR SRA in a SecDOAR platform has highlighted the need to **have an extendable security data orchestration approach** so that tools-specific data structures can be represented in the semantic model. In addition, we have also observed that for making an effective configuration of the SecDOAR platform, the security tools, metrics and visualisations need to be carefully selected so that a suite of tools can fill the gap in security data collection by gathering and analysing security data and visualising the results using appropriate security tools.

Instantiation of the SecDOAR SRA in a prototype using data from commercial tools (i.e., Snort, Splunk, Zeek and LimaCharlie) has identified certain challenges for implementing such systems. Firstly, if a large number of tools are used by a security operations team in an organisation, an *automatic approach to identify the relevant metrics required for analysis and for reporting* (e.g., visualisations) would be required. Secondly, the *selected dataset used for analysis should be comprehensive* so that it can provide all the parameters required for the security analysis metrics. If the data is not comprehensive, the frequency of false positives generated by the analysis can increase.

By presenting the SecDOAR SRA reported in this paper, we have attempted to fill the gap in the security data orchestration domain by contributing the meta-models, security data metrics with respect to the security operations, and a method to verify security data composition and details on SecDOAR SRA components that can lead to the design and development of the concrete systems for SecDOAR operations. In future, we plan to extend this research by focusing on runtime security tools composition and security tools integration optimisation so that security infrastructure can be provided following a Software as a Service (SaaS) model to the organisations. We plan to adapt and evaluate the SecDOAR SRA by incorporating machine learning models for security data analysis to see how these models can complement the security data metric for advanced analysis when using only the predefined metrics cannot provide conclusive results that can be used for cybersecurity-related decision making. In addition, we plan to investigate how security data labelling can be automated for the aforementioned purpose.

**Acknowledgments**

This research was supported by the Cyber Security Cooperative Research Centre Limited, whose activities are partially funded by the Australian Government's Cooperative Research Centre Program.